# A Cross Entropy test allows quantitative statistical comparison of t-SNE and UMAP representations

Carlos P. Roca[1*], Oliver T. Burton[1*], Julika Neumann[2,3], Samar Tareen[1], Carly E. Whyte[1], Stéphanie Humblet-Baron[2] and Adrian Liston[1]

[1]Immunology Programme, The Babraham Institute, Babraham Research Campus, Cambridge CB22 3AT, United Kingdom
[2]VIB Center for Brain and Disease Research, Leuven, Belgium.
[3]KU Leuven - University of Leuven, Department of Microbiology and Immunology, Leuven, Belgium.

*Correspondence should be addressed to adrian.liston@babraham.ac.uk

## Abstract

The advent of high dimensional single cell data in the biomedical sciences has necessitated the development of dimensionality-reduction tools. t-SNE and UMAP are the two most frequently used approaches, allowing clear visualisation of highly complex single cell datasets. Despite the ubiquity of these approaches and the clear need for quantitative comparison of single cell datasets, t-SNE and UMAP have largely remained data visualisation tools due to the lack of robust statistical approaches available. Here, we have derived a statistical test for evaluating the difference between dimensionality-reduced datasets, using the Kolmogorov-Smirnov test on the distributions of cross entropy of single cells within each dataset. As the approach uses the interrelationship of single cells for comparison, the resulting statistic is robust and capable of distinguishing between true biological variation and rotational symmetry generation during dimensionality reduction. Further, the test provides a valid distance between single cell datasets, allowing the organisation of multiple samples into a dendrogram for quantitative comparison of complex datasets. These results demonstrate the largely untapped potential of dimensionality-reduction tools for biomedical data analysis beyond visualisation.

**Introduction**

Single cell technologies are capable of generating vast datasets where multiple parameters (tens to thousands) are assessed on large numbers of cells (hundreds to tens of millions). With the rapid advance of these technologies, increasing the ease of data generation while decreasing the cost [1], data analysis approaches are rapidly becoming the limiting factor [2]. Low-dimensional visualisations are an attractive entry point to single cell analysis, with dimensionality-reduction tools providing an overview of the data. Beyond the science communication advantages of low-dimensional representation, such exploratory analyses often identify pattern distortions worthy of detailed analysis. Traditional approaches, such as Principle Components Analysis (PCA) and multidimensional scaling (MDS) produce linear-scaled low-dimensional representations, which are of high utility for comparing datapoints that are most different from each other. By contrast, the structure of most single cell experiments places the emphasis on identifying cells that are highly similar to each other; for these processes non-linear approaches provide superior resolution.

t-Distributed Stochastic Neighbor Embedding (t-SNE) is the most commonly used non-linear dimensionality reduction algorithm for single cell biology. In its common usage for visualising high-dimensionality single cell data, the algorithm starts with the single cells distributed at random points, along a Gaussian distribution, in transformed space. In an iterative process the cells move along a cost gradient, which provides a penalty for mismatch between the distances between two cells in the original high-dimensional space versus the representational low-dimensional space [3]. When sufficient iterations have occurred to reach stability, the outcome produces clusters of similar cells, based on the input data. Membership of a cluster indicates shared properties, however the non-linear nature of the penalty cost does not allow relationships to be inferred by the relative positioning of the clusters.

The wide-scale use of t-SNE to visualise single cell datasets has spurred the development of alternative non-linear dimensionality-reduction tools. Uniform manifold approximation and projection (UMAP) takes a similar iterative approach in positioning cells based on mismatch between high-dimensional and low-dimensional data, however it uses a distinct mathematical basis for the calculations. Cell positions are calculated in topological space, with mismatch between manifolds driving the iterative process [4]. While the results are broadly consistent with t-SNE visualisation, UMAP has superior run times and is argued to better preserve the global distances between cell clusters [4], although this is due to the choice of initialisation defaults rather than the UMAP algorithm [5]. den-SNE and densMAP are modifications of t-SNE and UMAP, respectively, which attempt to solve the problem inherent to these approaches whereby cluster size is driven by the number of cells in the cluster. den-SNE and densMAP use an auxiliary function to optimise cell density between high-dimensional and low-dimensional space, with the size of clusters in the resulting low-dimensional plots reflecting the degree of heterogeneity present in the cluster [6]. Additional non-linear tools have been built, such as TMAP [7], to deal with limitations of t-SNE when applied beyond the single cell space.

The near-ubiquity of t-SNE or analogous tools in the visualisation of single cell sequencing data demonstrates the utility of low-dimensional representations of complex data. Despite this, t-SNE is generally not used as an analytical tool, with downstream analysis most commonly based on subsetting data and treating as pseudo-bulk populations. Here we have developed a Cross Entropy test allowing robust statistical comparisons of t-SNE representations. The test has appropriate sensitivity, with the identification of biological differences and the dismissal of difference in technical replicates, biological replicates or repeat t-SNE runs. The test responds to differences in either inter-cluster frequency or intra-cluster phenotype shifts, and differences between multiple t-SNE plots can be quantified for the production of dendrograms. The Cross Entropy test is broadly applicable to single cell technologies, including flow cytometry, mass cytometry and single cell sequencing, and can be performed on either t-SNE and UMAP transformations, providing a highly versatile statistical tool to the single cell toolkit.

**Results**

*A Cross Entropy test provides a robust statistical test for t-SNE comparison*

t-SNE is an iterative algorithm for dimensionality reduction. In its common usage for visualising high-dimensionality single cell data, the cost gradient of t-SNE places greater weight on pairs of cells close to each other, with medium- and long-range pairs ignored. The low-dimensional representation of high-dimensional data makes t-SNE an attractive visualisation tool, yet its value as an analytical tool has been hampered by the paucity of statistical tests. We sought to derive a statistical test capable of distinguishing biological differences in single cell t-SNE representations, while being robust against false detection of differences in technical replicates or the seed-dependent variation in t-SNE generation. As the t-SNE algorithm is driven by the cross entropy of the individual cells in the dataset, and the t-SNE fixes the average point entropy, each t-SNE can be considered a distribution of cross entropy divergences. Deriving a distribution of cross entropy divergences per t-SNE plot therefore allows the use of the Kolmogorov-Smirnov (KS) test to evaluate the degree of difference between two, or more, t-SNE plots (see Methods).

A robust t-SNE statistical test should: 1) reliably fail to detect differences when comparing technical replicates; 2) reliably fail to detect differences when comparing biological replicates; 3) reliably identify differences when comparing different biological samples; and 4) not get fooled by different t-SNE runs on the same sample. The Cross Entropy test was beta-tested by collaborating immunologists on >130 datasets over the course of 24 months of validation. We initially focused on flow cytometry datasets, as the very high cell number achievable by flow cytometry provides superior ability to challenge and validate a statistical test. As an example of the robustness observed, we present here a high-dimensional flow cytometry dataset based on immunological profiling of lymphocytes from the lymph nodes, spleen and tissues from the C57BL/6 inbred mouse strain (MUS dataset). This design allows us to compare technical replicates (splitting a single sample), biological replicates (comparing analogous samples from different mice) and biological samples (comparing lymphocytes from different tissues of the same mice). First, to test for robustness in technical replicates, we compared the t-SNE plots generated by splitting a single sample of splenocytes, with FlowSOM clustering allowing comparative visualisation of cell identity (**Figure 1A**). Visually similar plots were generated, with Cross Entropy test values showing p values of 0.370 to 1, supporting the null hypothesis of no difference (**Figure 1A**). Second, to test for robustness in biological replicates, we compared the t-SNE plots generated by multiple samples of splenocytes from age-/sex-matched mice. Again, visually similar plots were generated, with the Cross Entropy test p values of 0.202 to 0.636 supporting the null hypothesis (**Figure 1B**). Third, to test for the capacity to identify true biological differences, we compared the t-SNE plots generated on lymphocytes profiled from the spleen, lymph node and tissues of mice. Lymphocytes from the lymph nodes and spleen are phenotypically similar, but the relative proportions of each population vary between the sites, while non-lymphoid tissue lymphocytes are phenotypically distinct. t-SNE visualisations mirrored this biological background knowledge, with highly significant Cross Entropy test values across each comparison (**Figure 1C**). Fourth, we produced independent runs of t-SNE generation on the lymph node sample, generating similar t-SNE plots with the characteristic rotational symmetry (**Figure 1D**). Despite the visual disparity being greater than the highly significant spleen-lymph node comparison, the Cross Entropy test gave a test p value of 0.585, supporting the null hypothesis (**Figure 1D**). Multiple runs of this false comparison gave an appropriate sensitivity of the Cross Entropy test, with neither over- nor under-reporting of false positives (**Figure 1E**). Together, this example dataset demonstrates the robust ability of the Cross Entropy test to distinguish biological signal from noise.

*Calculation of relative distance between t-SNE comparisons*

A key advantage of the utilisation of the Kolmogorov-Smirnov test in the Cross Entropy test is the ability to calculate the $L^\infty$ distance, which can be used as a measure of the difference between t-SNE plots. In order to test the ability of the $L^\infty$ distance to correctly assign distances between t-SNE plots, we created two artificial datasets of known distance from the splenic and tissue flow cytometry data. First, we created a spleen$^{tissue}$ dataset, where 90% of the events were extracted from the spleen sample and 10% of events were extracted from the tissue sample prior to concatenation to create a single merged sample. Second, we created the reciprocal tissue$^{spleen}$ dataset, with 90% tissue events and a 10% splenic event spike-in. Generation of t-SNE plots using the parental samples and constructed spike-in samples provides visual confirmation of the intermediate status of the artificial datasets (**Figure 2A**). Calculation of the $L^\infty$ distance allowed the construction of a dendrogram that correctly assigned spleen$^{tissue}$ as biologically closer to spleen, and tissue$^{spleen}$ as biologically closer to tissue (**Figure 2B**). This demonstrates the $L^\infty$ distance as a tool for gauging the relative closeness of t-SNE plots when comparing multiple samples.

*A Cross Entropy test responds to both quantitative and qualitative changes in single cell phenotype*

Next we sought to determine the sensitivity of the Cross Entropy test to biological differences based on inter-cluster frequency versus intra-cluster phenotype. Currently, most single cell analysis ultimately measures either the frequency of individual clusters (typically annotated based on FlowSOM [8] or similar clustering tools [9]), or treats individual clusters as pseudo-bulk populations for comparison of gene/protein expression changes. In principle, as t-SNE incorporates the full phenotype of individual cells, a useful t-SNE-based statistical test should be sensitive to both changes in inter-cluster frequency and changes in intra-cluster phenotype. To test this, we generated artificial datasets derived from the MUS dataset of lymphocytes from the spleen and lymph node, which exhibit changes in both inter-cluster frequency and phenotype. First, we used selective downscaling to create a lymph$^{\%spleen}$ dataset, where the number of cells in each major FlowSOM cluster were normalised to match those in the spleen dataset, while only using cells from the lymph node data. Comparison of the spleen sample to the re-scaled lymph$^{\%spleen}$ sample generated t-SNE plots with cluster frequencies normalised between spleen and LN (**Figure 3A**). Despite this, the Cross Entropy test identified the samples as significantly different (**Figure 3B**), based on the phenotype difference within clusters (**Figure 3C**). Next, we created a reciprocal artificial, spleen$^{\%lymph}$, taking biological replicates of the spleen sample and selectively downscaling populations to create cluster frequencies similar to that of the LN (**Figure 3D**). To determine whether the Cross Entropy test could detect differences solely based on cluster frequency, we compared spleen to spleen$^{\%lymph}$. The test found significant differences between the two samples (**Figure 3E**), despite identical phenotypes (**Figure 3F**). These results demonstrate that the Cross Entropy test is sensitive to both qualitative and quantitative changes between clusters.

*The Cross Entropy test has broad utility in comparison of dimensionally-reduced single cell datasets*

Having demonstrated the validity of the Cross Entropy test on a mouse-based high dimensional flow cytometry panel, we sought to test it on the independent single cell technologies of mass cytometry and single cell sequencing from human samples. We used available datasets generated from healthy humans and COVID-19 patients, based on mass cytometric analysis of the peripheral blood [10], and 10x single cell sequencing analysis of bronchoalveolar lavage [11]. Using the mass cytometry dataset, we compared peripheral blood lymphocyte subsets from 12 patients at different time points, namely admission to the ICU, during their stay at the ICU (intermediate) and upon discharge from the ICU. We identified major subsets based on characteristic marker expression, with the dataset reproducing key features of severe COVID-19, such as increased neutrophils and decreased $CD4^+$ as well as $CD8^+$ T cells [12, 13] (**Figure 4A**). The $L^\infty$ distance between the three time points suggests that the overall immune landscape during the ICU stay more closely resembles that of the admission time point (**Figure 4A**). However, when performing analysis of the monocyte subset alone, we observed a closer resemblance of the intermediate time point to that of discharge (**Figure 4B**). These results are in accordance with findings from the original publication, generated through traditional gating and marker analysis, as well as other studies that report monocytes as the first immune population to recover following severe COVID-19 [10, 14]. Turning to the single cell sequencing dataset, we used a 10x single cell sequencing comparison of bronchoalveolar lavage from COVID patients and non-COVID pneumonia patients [11]. Using the annotated cell clusters, we compared the transcriptional profile of COVID to non-COVID pneumonia cells for each of the epithelial, neutrophil, monocyte/macrophage, CD4 T cell, CD8 T cell, dendritic cell, B cell and NK cell clusters (**Figure 4C**). Highly significant changes were observed in the t-SNE cross entropy of COVID vs non-COVID epithelial, neutrophil, monocyte/macrophage, CD4 T cell and CD8 T cell, in accordance with the detailed classical analysis used [11]. Use of the $L^\infty$ distance (**Figure 4D**) identified the largest change being in the neutrophil compartment, consistent with the results from multiple studies [11, 15, 16], and found the degree of change in CD8 T cells to be greater than that observed in CD4 T cells, again consistent with traditional analysis [11]. Together these results demonstrate that the Cross Entropy test is compatible with t-SNE analysis generated by multiple independent technologies, with a simple test recapitulating many of the key features identifying through high-depth traditional methods.

*The Cross Entropy test has broad utility in comparison of dimensionally-reduced single cell datasets*

Recently, alternative nonlinear data visualisation tools have been generated for single cell analysis. UMAP performs the same basic functions as t-SNE with regards to using two dimensions to separate cells based on complex multi-dimensional phenotype data [4]. Despite the different mathematical basis for generating the dimensionality reduction, cross entropy dictates the separation of UMAP as well as t-SNE. We therefore performed validation tests for the Cross Entropy test on UMAP representations. Using the MUS dataset, we produced UMAP plots comparing technical replicates (**Figure 5A**), biological replicates (**Figure 5B**), biological differences (**Figure 5C**) and repeat UMAP runs (**Figure 5D**). In each case, the Cross Entropy test produced the appropriate conclusions, failing to detect significant differences between technical and biological replicates, while appropriately detecting differences in biologically-distinct samples. The Cross Entropy test was also appropriately powered, with neither over- nor under-sensitivity detected (**Figure 5E**). As with the utility in the t-SNE test, the UMAP Cross Entropy test respond to both shifts in sub-cluster cellular phenotype and cluster frequency, using the artificial samples Spleen$^{\%LN}$ and LN$^{\%spleen}$, compared to the parental spleen sample (**Figure 5F**). These results demonstrate the broad utility of the Cross Entropy test for multi-dimensionally scaled single cell data, across both t-SNE and UMAP approaches.

**Discussion**

The advent of single cell sequencing has led to an urgent need for analysing and visualising the complex high-dimensional datasets generated. While similar utility can be found in flow and mass cytometry datasets, the lower number of parameters and increased reliability of individual parameters allowed the continuation of the pairwise gating approach, derived from the historical progress of biological knowledge and technical capacity [9]. Both t-SNE and UMAP approaches fulfil the visualisation role created by single cell technologies, especially when paired with clustering tools such as FlowSOM [8], to provide biological meaning to the two-dimensional space that represents complex shifts in phenotype or transcriptome. At an analytical level, however, t-SNE and UMAP have been under-utilised. Most analytical approaches treat single cell data as pseudo-bulk data. Single cell data is typically only used to generate clusters as an intermediate step [9]. Downstream analysis typically compares cluster size change between samples, or compares cluster-aggregated data for cluster-level statistics (such as mean fluorescence intensity in flow cytometry or average transcript expression in single cell sequencing). This large information loss provides an opportunity for the generation of new statistical tests that include single cell information in the sample comparisons.

Unlike the commonly-used pseudo-bulk approach, t-SNE and UMAP encapsulate the nuance of data at the single cell level. Cross entropy captures, as a single statistic, both changes in the relative frequency of cells of different phenotype classes and also the shift of phenotype of cells within phenotype classes. We have used this tool to quantify the complex shifts in phenotypic markers present in T cells of different regions of brain, or transcriptional shifts occurring in microglia in the presence or absence of CD4 T cells [17]. Valid targets for the Cross Entropy test include any high-dimensional single cell dataset, routinely generated by single cell sequencing, flow cytometry or mass cytometry in fields ranging from immunology [18] to neuroscience [19] to cancer [20]. The growth of single cell technologies expands the potential range to techniques such as CODEX or other high-dimensional imaging technology [21], single cell proteomics [22] or mutation analysis in single cell sequencing [23]. More broadly, t-SNE and UMAP use can be implanted in cases where the cell is not the individual unit. For example, using people as the individual unit, high-dimensional data such as immune response [24], genomic variation [25], or microbiome composition [26], can drive the algorithm. The Cross Entropy test would then allow statistical comparison of disparate groups (e.g., based on disease state) to identify differences in the underlying data. In each case, full utility of the data generated requires a statistical tool that can compare the high-dimensional data distributions between two groups.

The use of the Kolmogorov-Smirnov tests, and the ability to quantify differences between samples, in the Cross Entropy test provides a potential route to incorporate t-SNE or UMAP into routine diagnostics. Flow cytometric analysis of blood leukocytes is routinely used as a single cell technology in the clinic [27], however data analysis is largely limited to calculate of population frequencies compared to reference populations. Consortia such as EuroFlow are standardising protocols to enable a broader uptake and cross-centre comparison [28]. The use of the cross entropy $L^\infty$ distance would enable a simple analysis for deviations in leukocyte number or activation status, by determining whether an individual sample mapped to the healthy or disease nodes of a t-SNE-based dendrogram. For example, we used Cross Entropy to draw a dendrogram of T cell phenotypes in individuals with mild or severe COVID [29]. Broader use of a Cross Entropy test approach within the diagnostics field may provide higher sensitivity to the detection of atypical immunological disorders or haematological malignancies, as the integrated statistical would incorporate the detection of both aberrant marker expression and the development of rare/unusual populations not routinely gated for [30]. While flow cytometry is the currently the dominant single cell technology in diagnostics, both mass cytometry [31] and single cell sequencing [32] are actively being developed for diagnostics and may soon become routine; parallel development of the Cross Entropy statistic may enhance the utility and sensitivity of these new technologies.

Despite the high utility of the Cross Entropy test for calculating the significance of differences between two or more dimensionality-reduced plots, we caution against the misuse of a p value as a measure of biological meaningfulness. The power of Cross Entropy tests is dependent on cell number, and thus even biologically-distinct samples can return non-significant p values if the number of cells is highly limiting. In practice, sample sizes in the low thousands range can return non-significant p values if the biological difference is slight, while datasets above 10,000 cells are highly powered for even sensitive change. Conversely, very high cell numbers can provide sufficient power to capture the differences between biological replicates, which, while real, are rarely considered to be biologically meaningful. In the datasets tested here, for high cell number analysis a threshold of 0.001 excludes biological replicates while successfully identifying biological differences. We suggest that analyses include biologically-distinct positive controls and the use of the $L^\infty$ distance, in addition to the p value calculation. The $L^\infty$ distance, as a measure of difference magnitude rather than statistical difference, is less sensitive to cell number changes, and the use of a biologically-distinct positive control provides a reference point for the magnitude of the differences observed. Finally, as with all statistical tests, the utility of the analysis is fully dependent on the quality of the data and experimental design.

**Methods**

*Datasets*

Collaborating immunologists beta-tested the Cross Entropy test, testing for robustness. Among these datasets, four are used as examples here, covering high dimensional flow cytometry in mice (MUS), and flow cytometry (FC), mass cytometry (MC) and single cell sequencing (SCS) datasets in humans.

The MUS dataset may be accessed via FlowRepository (https://flowrepository.org/id/FR-FCM-Z48W). For the MUS dataset, spleen, lymph nodes and small intestinal lamina propria from 10-week old female C57Bl/6 mice were stained for flow cytometry. Spleen and lymph nodes were disrupted with glass slides, filtered through 100 μm mesh, and, in the case of the spleen, red blood cells were lysed. Intestinal tissue was incubated in staining buffer (HBSS with 10mM HEPES, 2% FCS and 2mM EDTA) (Gibco) for 15min at 37°C in a shaking platform. Dislodged cells were discarded and the remaining tissue was digested with 1mg/ml Collagenase D (Roche), 100μg/ml hyaluronidase and 40μg/ml DNAse I (Sigma) in IMDM with 10mM HEPES and 10% FCS) for a further 30min at 37°C with shaking. The tissue was then filtered through 100 μm mesh to obtain a single cell suspension. The intestinal single cell suspension was passed through a 40% Percoll (GE Healthcare) gradient by centrifugation at 600g for 10min at 4°C. Cells ($2\times10^6$) were blocked for 30min with 2.4G2 hybridoma supernatant supplemented with 1% Monocyte Blocker (BioLegend) prior to staining. Cells were stained for 60min at 37°C with CD45-NovaBlue 530, CD4-NovaBlue 585, CD8-NovaBlue 610, CD19-NovaYellow 690, CD3-NovaRed 685, CD69-biotin (ThermoFisher), CCR2-BV750 and CCR9-BB515 (BD Biosciences) in staining buffer containing presence of CellBlox (NovaBlock). Cells were washed and stained with ViaKrome 808 (Beckman Coulter) and Qdot705-streptavidin (ThermoFisher) in HBSS at room temperature for 20min. After washing, cells were fixed and permeabilized with Foxp3 Transcription Factor Staining Buffer Set (eBioscience) according to the manufacturer's instructions. Cells were then washed twice with permeabilization buffer (eBioscience) and stained overnight (16hrs) at 4°C in permeabilization buffer with 2.4G2 anti-CD16/32 for the following antibodies: CD103-BUV395, IgD-BUV496, NK1.1-BUV563, CTLA-4-BUV615, c-Kit-BUV661, CD62L-BUV737, GITR-BUV805, CXCR3-BV480, Siglec F-BV510, TCRγδ-BB660-P2, PDCA-1-BB700, CD11c-BB755-P, Ly-6C-BB790-P (BD Biosciences), CD150-BV421, Helios-Pacific Blue, TCRβ-BV570, PD-1-BV605, XCR1-BV650, CD127-BV711, I-A/I-E-PerCP, CD64-PE-Dazzle 594, CD11b-PE-Fire 640, CD172a-PE-Cy7, CD38-PE-Fire 810, CD44-APC-Fire 750, Gr-1-APC-Fire 810 (BioLegend), ICOS-Super Bright 436, KLRG1-Super Bright 780, CD86-PE-Cy5, NKp46-PerCP-eFluor710, F4/80-AF561, Foxp3-PE-Cy5.5, RORγT-APC, GATA-3-PE, T-bet-eFluor660, Ki67-AF700 (Thermo Fisher), B220 (CD45R)-StarBright UV 445, CD25-StarBright Violet 515 (Bio-Rad), CD24-Pacific Orange, IgM-AF532 and CD90.2-AF790 (conjugated in house). Cells were washed twice with permeabilization buffer and once with staining buffer prior to acquisition. Samples were acquired on a 5-laser Aurora Spectral Analyser (Cytek), with unmixing based on single stained cell controls. An autofluorescence channel was created using unstained intestinal cells, gating on high SSC cells.

For the artificial MUS dataset, new FCS files were created using events extracted and concatenated from the MUS dataset. For lymph-spleen normalization, clusters mapped by FlowSOM were exported as independent FCS files, downscaled in FlowJo to match the new proportions, and concatenated. For tissue-spleen spike-in data, events were extracted and concatenated in R using flowCore and premessa.

For the mass cytometry dataset, we used data published by Pentilla et al.[10] and available in the Flow repository (https://flowrepository.org/id/FR-FCM-Z34U). We selected 12 individuals that were serially sampled at the time of admission to the ICU, 6-8 days after (intermediate time point) and upon discharge from the ICU. Briefly, whole blood was collected and stained with antibodies outlined in the original publication within 2-4 h of collection and acquired on a Helios® Mass Cytometer (Fluidigm). For the analysis performed in this manuscript, living single cells were gated and beads removed. Monocytes were identified as being $CD45^+CD3^-CD19^-$ $CD14^+$ or $CD16^+$ and $CD56^-$.

For the SCS dataset, we used data published by Wauters et al[11] and available in the EGA European Genome-Phenome Archive database (EGAS00001004717). Briefly, bronchoalveolar lavage was collected from 22 COVID-19 patients and 13 non-COVID-19 pneumonia patients during standard-of-care treatment. The cellular component of the lavage was used for single cell sequencing on the 10x platform. We used the data from the bronchoalveolar lavage of both COVID and non-COVID pneumonia patients, with the published cell identity. For the presented analysis, t-SNE representations were calculated based on the count matrices using the Seurat pipeline and its coordinates used to calculate the cross entropy distributions and comparison.

*t-SNE representation of multidimensional data*

The t-SNE algorithm aims to provide a simplified, low-dimensional representation of high-dimensional data, while preserving as much as possible the local distances between data points, that is, the differences between similar data points.

Let us represent a dataset in original space as a collection of n data points $\{x_i\}$, with $i = 1 \ldots n$. All data points $x_i$ are vectors of $\mathbf{R}^d$, with $d$, the number of dimensions, being usually a large number.

The low-dimensional representation that t-SNE produces is another collection of data points $\{y_i\}$, again with $i = 1 \ldots n$. In this case, data points are vectors of a real space with low number of dimensions, typically $\mathbf{R}^2$ or $\mathbf{R}^3$.

What defines the t-SNE algorithm, as well as any other technique of dimensionality reduction, is the way in which the new collection of points $\{y_i\}$ is obtained from the original collection $\{x_i\}$. In the case of t-SNE, Gaussian probabilities are defined for each pair of points, as follows. Given a pair in the original space $(x_i, x_j)$, with $i \neq j$, the probability $p_{ij}$ is defined by assigning first to each tuple $(i,j)$:

Equation 1

$$p_{j|i} = \frac{exp(-\|x_j - x_i\|2\sigma_i^2)}{\sum_{k \neq i} exp(-\|x_k - x_i\|2\sigma_i^2)}$$

and then symmetrizing by

Equation 2

$$p_{ij} = \frac{p_{i|j} + p_{j|i}}{2n}$$

which implies the global normalization property

Equation 3

$$\sum_{i=1}^{n} \sum_{\substack{j=1 \\ j \neq i}}^{n} p_{ij} = 1$$

The parameters $\sigma_i$ are adjusted for each point $x_i$, so that the perplexity $\rho$ (or equivalently the entropy $\log \rho$) of the distribution $\{p_{j|i}\}$ has a predefined value.

Similarly, probabilities in the representation space are defined for each pair of points $(y_i, y_j)$, with $i \neq j$, but following in this case a Cauchy distribution:

Equation 4

$$q_{ji} = \frac{(1 + \|y_j - y_i\|^2)^{-1}}{\sum_{k \neq i} (1 + \|y_l - y_k\|^2)^{-1}}$$

which implies global normalization without the need of symmetrisation

Equation 5

$$\sum_{i=1}^{n} \sum_{\substack{j=1 \\ j \neq i}}^{n} q_{ij} = 1$$

With the definition of these probability distributions, which give more weight to pairs of points close to each other, the problem of finding the transformed points $\{y_i\}$ from the original points $\{x_i\}$ is converted into the problem of making the probabilities $\{q_{ij}\}$ as similar as possible to $\{p_{ij}\}$. This is achieved by minimizing numerically the Kullback-Leibler divergence between the distributions $P = P = \{p_{ij}\}$ and $Q = \{q_{ij}\}$,

Equation 6

$$D(P,Q) = \sum_{i=1}^{n} \sum_{\substack{j=1 \\ j \neq i}}^{n} p_{ij} \log \frac{p_{ij}}{q_{ij}}$$

This is equivalent to making the cross entropy between P and Q

Equation 7

$$H(P,Q) = -\sum_{i=1}^{n} \sum_{\substack{j=1 \\ j \neq i}}^{n} p_{ij} \log q_{ij}$$

as close as possible to the entropy of P

Equation 8
$$H(P) = - \sum_{i=1}^{n} \sum_{j=1, j \neq i}^{n} p_{ij} \log p_{ij}$$

given that

Equation 9
$$D(P,Q) = H(P,Q) - H(P)$$

Because the entropy of the distribution *P* is fixed by construction to a value determined by the perplexity $\rho$, the optimization carried out by t-SNE only operates over the cross entropy between P and Q.

*Distributions of entropy and cross entropy*

The key insight for the test on t-SNE representations proposed here resides in considering the distributions of entropy and cross entropy per point, instead of their global counterparts. For a data point with index *i*, the entropy and cross entropy, $\{H_i^{(P)}\}$ and $\{H_i^{(P,Q)}\}$, respectively, are defined as:

Equation 10
$$H_i^{(P)} = - \sum_{j=1, j \neq i}^{n} p_{ij}^* \log p_{ij}^*$$

Equation 11
$$H_i^{(P,Q)} = - \sum_{j=1, j \neq i}^{n} p_{ij}^* \log q_{ij}^*$$

with

Equation 12
$$p_{ij}^* = n p_{ij}$$

Equation 13
$$q_{ij}^* = n q_{ij}$$

The local pair probabilities $p_{ij}^*$ and $q_{ij}^*$ are introduced to have quantities with a sum per point closer to 1:

Equation 14
$$\sum_{j=1, j \neq i}^{n} p_{ij}^* \approx 1$$

Equation 15

$$\sum_{j=1, j\neq i}^{n} q_{ij}^* \approx 1$$

The distribution of point Kullback-Leibler divergences can also be obtained as

Equation 16

$$D_i^{(P,Q)} = H_i^{(P,Q)} - H_i^{(P)}$$

This way, the global Kullback-Leibler divergence can be written as the average of the distribution of divergences per point

Equation 17

$$D(P,Q) = \frac{1}{n}\sum_{i=1}^{n} D_i^{(P,Q)} = \frac{1}{n}\sum_{i=1}^{n} H_i^{(P,Q)} - \frac{1}{n}\sum_{i=1}^{n} H_i^{(P)}$$

By design, t-SNE makes all the point entropies $\{H_i^{(P)}\}$ close to a fixed value determined by the perplexity $\rho$, with $H_i^{(P)} = \log\log\rho$ for all $i$ before symmetrization. Therefore, their average is fixed a priori and almost the same for any input dataset $\{x_i\}$. As a consequence, differences between results obtained from different input datasets will necessarily appear as differences between the respective distributions of cross entropy per point $\{H_i^{(P,Q)}\}$.

*Test on distributions of cross entropy*

Given two input datasets $\{x_i\}$ and $\{x_i'\}$, the t-SNE algorithm will produce two representations $\{y_i\}$ and $\{y_i'\}$, with associated distributions of cross entropy $\{h_i\}$ and $\{h_i'\}$.

Under the null hypothesis, $\{x_i\}$ and $\{x_i'\}$ will follow the same probability distribution, which implies that local distances between pair of points will also be equal in distribution. As a consequence, after applying the t-SNE algorithm, obtained datasets in representation space $\{y_i\}$ and $\{y_i'\}$, local distances, and point cross entropies will also be equal in distribution.

Under the alternative hypothesis, $\{x_i\}$ and $\{x_i'\}$ will differ in probability distribution. This will likely imply differences in the distributions of local distances, which will in turn produce differences in distribution for the local distances in representation space. As a result, point cross entropies will likely differ in distribution.

Therefore, the test consists in evaluating the difference between the distributions of cross entropy, which allows us to use already available tests for comparing one-dimensional empirical distributions. The main test chosen for this purpose is the Kolmogorov-Smirnov test, which offers the additional advantage of providing a statistic that is a valid distance between functions, namely the $L^\infty$ distance. This distance can be used as a measure of the difference between t-SNE plots, for example to organize them into a dendrogram

*General implementation details*

The Cross Entropy test was implemented in R v.4.0.5, using the packages ConsensusClusterPlus, digest, dunn.test, flowCore, FlowSOM, ggplot2, ggridges, RANN, RColorBrewer, reshape2, Rtsne and umap. The script used is available on GitHub at https://github.com/AdrianListon/Cross-Entropy-test. Parallel scripts are available for use on cytometry and single cell sequencing datasets. As a practical guide for biologists with low experience at working in R, we have provided a walk-through on the use of this code at: https://www.liston.babraham.ac.uk/flowcytoscript/

**Acknowledgments**

This project has received funding from the European Union's Horizon 2020 research and innovation programme under grant agreement No 779295. This work was also supported by the ERC Consolidator Grant TissueTreg (to AL), the Biotechnology and Biological Sciences Research Council through BB/S019189/1, Institute Strategic Program Grant funding BBS/E/B/000C0427 and BBS/E/B/000C0428, and the Biotechnology and Biological Sciences Research Council Core Capability Grant to the Babraham Institute. The authors thanks the Babraham Institute bioinformatics and flow cytometry core facilities.

**Author Contributions**

The study was conceived by CPR and AL. CPR developed and tested the Cross Entropy test. OTB, CW, JN, ST and SHB provided the datasets and evaluated results. The manuscript was written by CPR and AL, and revised and approved by all authors.

# Figures

**Figure 1. A Cross Entropy test provides a robust statistical test for t-SNE comparison.**
The MUS flow cytometry dataset was used to test the sensitivity of the cross entropy test. **A)** Three technical replicates of a single splenocyte sample were plotted on a t-SNE with FlowSOM clustering in overlay. P values for Cross Entropy testing are shown. **B)** t-SNE plots of splenocytes from three individual age/sex-matched mice (biological replicates). P values for Cross Entropy testing are shown. **C)** t-SNE plots from lymph nodes, spleen and tissue (small intestinal lamina propria). P values for Cross Entropy testing are shown. **D)** Two independent t-SNE runs of lymph node cells. P values for Cross Entropy testing are shown. **E)** Cumulative distribution function of p-values obtained from 200 comparisons of independent t-SNE runs of the same lymph node sample. Statistical comparison to uniform distribution using KS test.

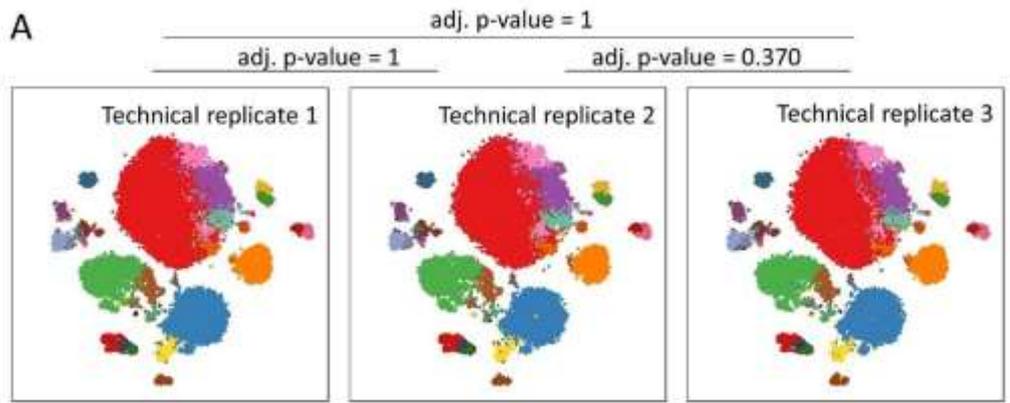
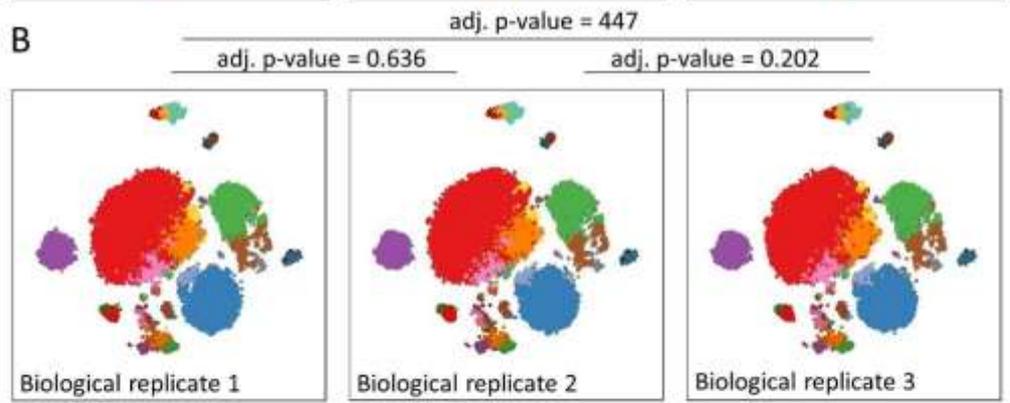
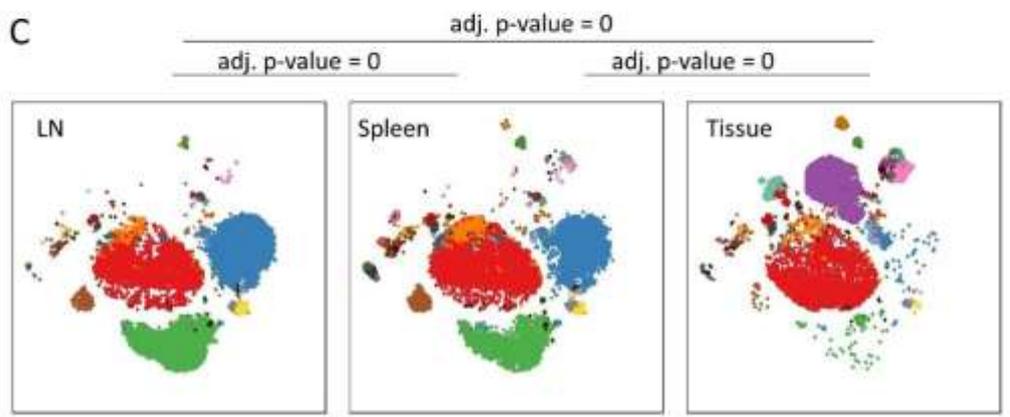
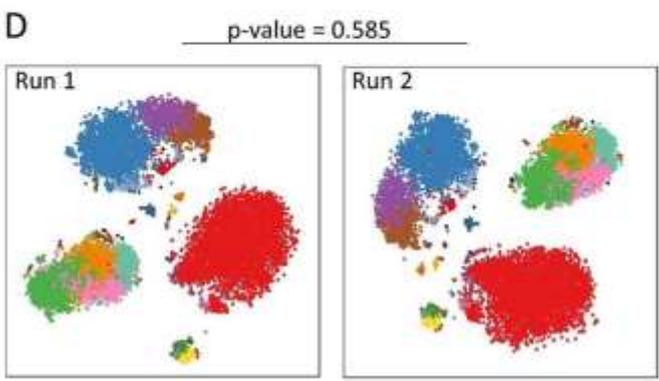
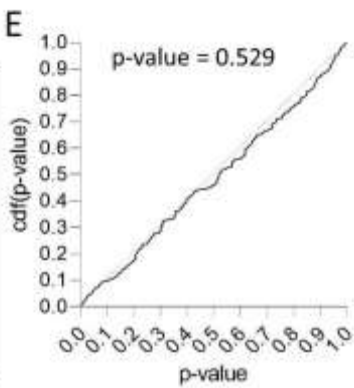

**Figure 2.** $L^\infty$ **provides a quantitative comparison of different t-SNE visualisations.**
In order to create a series of samples with defined degrees of biological difference, we started with the spleen and tissue (small intestinal lamina propria) samples from the MUS sample set and generated two artificial samples, one composed of 90% spleen and 10% tissue cells (spleen$^{tissue}$), and one composed of 10% spleen and 90% tissue cells (tissue$^{spleen}$). **A)** t-SNE plots showing spleen, spleen$^{tissue}$, tissue$^{spleen}$ and tissue datasets. **B)** Dendrogram based on the $L^\infty$ cross entropy distance.

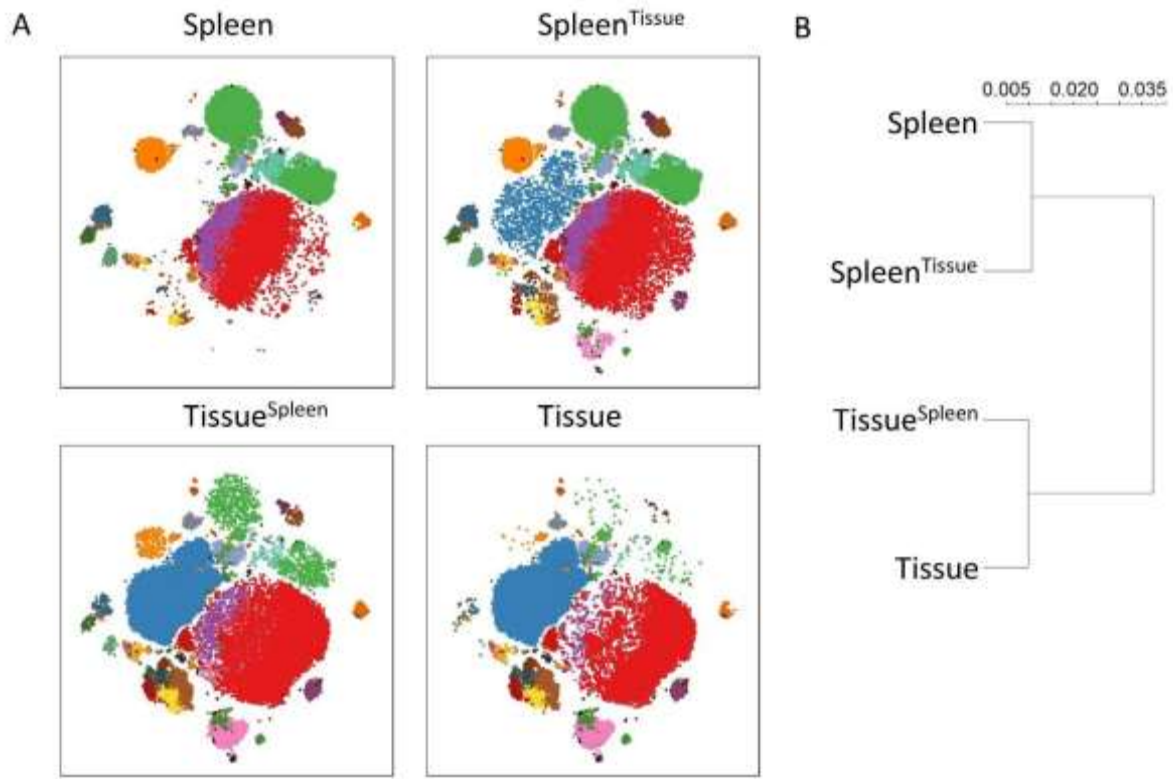

**Figure 3. Cross entropy test detects both qualitative and quantitative changes in single cell datasets.** In order to create a series of samples with biological difference restricted only to qualitative or quantitative changes, we started with the spleen and lymph node samples from the MUS sample set and generated artificial samples. Artificial samples were generated by first defining and exporting the clusters in each parental sample, then downscaling to match the new proportions and concatenating to reconstruct the artificial sample. This resulted in a sample where lymph node clusters were rescaled, such that each cluster was the same frequency as the analogous cluster in spleen ($LN^{\%spleen}$) and the reciprocal sample where spleen clusters were rescaled to the same frequency as the analogous cluster in lymph nodes ($Spleen^{\%LN}$). **A)** FlowSOM cluster frequencies in spleen and $LN^{\%spleen}$ samples. **B)** t-SNE plots of in spleen and $LN^{\%spleen}$ samples. P values for Cross Entropy testing are shown. **C)** Bar chart showing selected MFI values per cluster for spleen and $LN^{\%spleen}$ samples. **D)** FlowSOM cluster frequencies in spleen, $Spleen^{\%LN}$ and lymph node samples. **E)** t-SNE plots of $Spleen^{\%LN}$ and spleen samples. P values for Cross Entropy testing are shown. **F)** Bar chart showing MFI values for markers in clusters from $Spleen^{\%LN}$ and spleen samples.

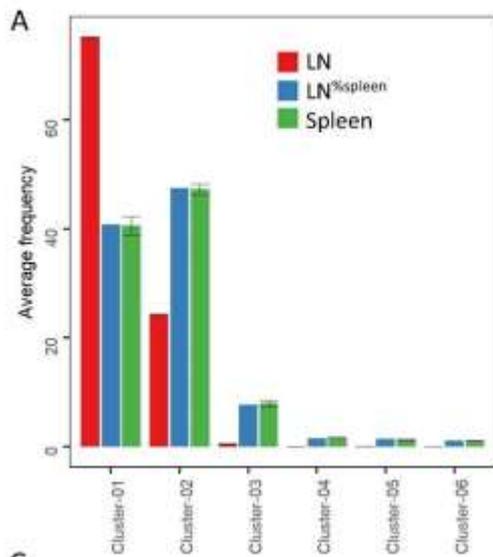
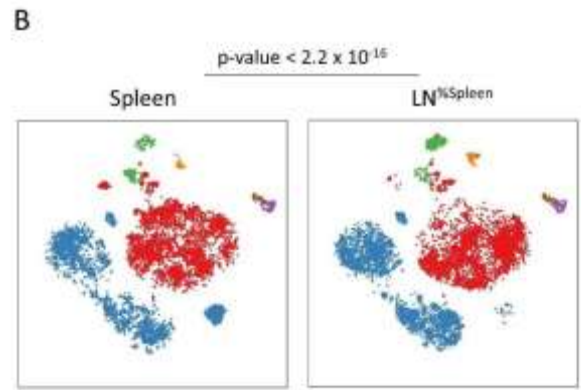
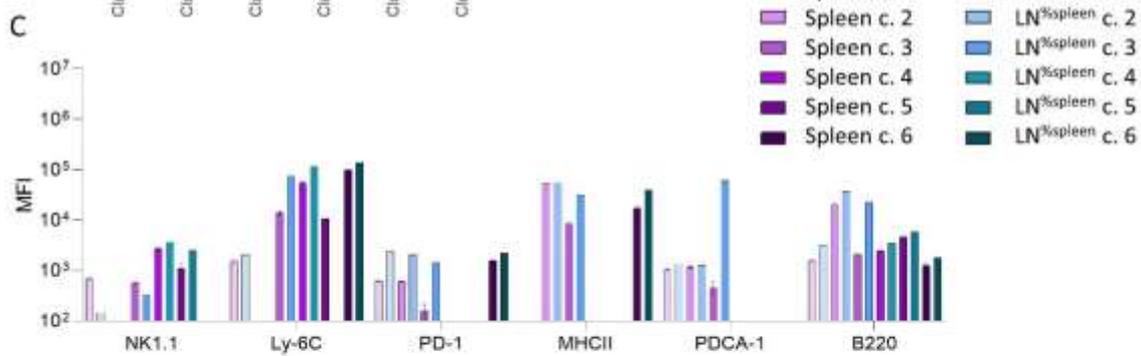
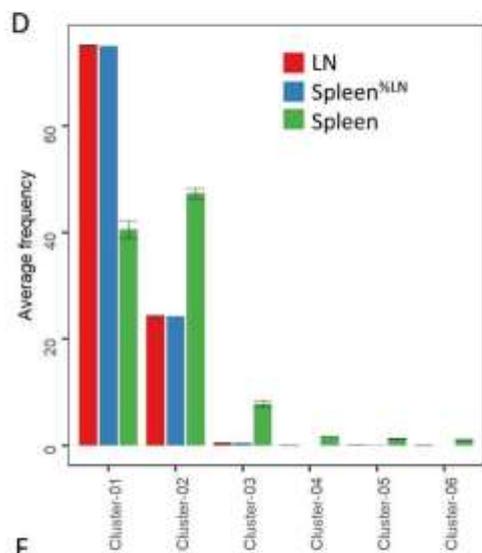
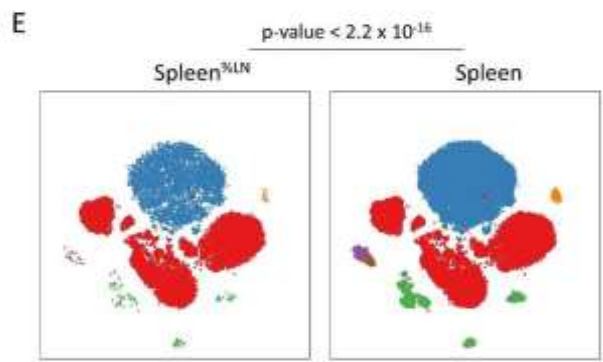
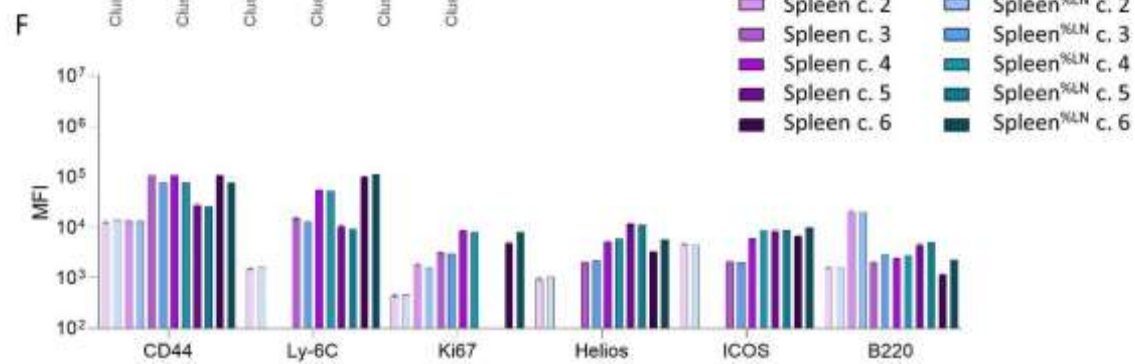

**Figure 4. Cross Entropy test provides utility on mass cytometry and single cell sequencing datasets. A)** Whole blood from COVID-19 patients sampled upon admission, during their stay (intermediate) and upon discharge from the ICU was analyzed by CyTOF. Total cells were clustered using the t-SNE algorithm with FlowSOM clusters overlaid and annotated (left). Bar chart represents the mean frequency of each cluster per condition (middle). Dendrogram showing the comparative similarity between conditions based on the KS statistic (right). **B)** t-SNE run on monocytes from COVID-19 patients as in A, overlaid with FlowSOM clusters (left) and events per condition (middle). Dendrogram showing the comparative similarity between conditions based on the KS statistic (left). **C)** Bronchoalveolar lavage from COVID patients and non-COVID pneumonia patients was assessed by 10x single cell sequencing. Cells were annotated as epithelium, neutrophils, monocytes/macrophages. CD4 T cells, CD8 T cells, dendritic cells, B cells and NK cells, based on transcriptional profiles. For each cell type, cells were clustered using the t-SNE algorithm with patient source overlaid. Cross entropy p value and **D)** $L^\infty$ distance between COVID and non-COVID patients for each cell type.

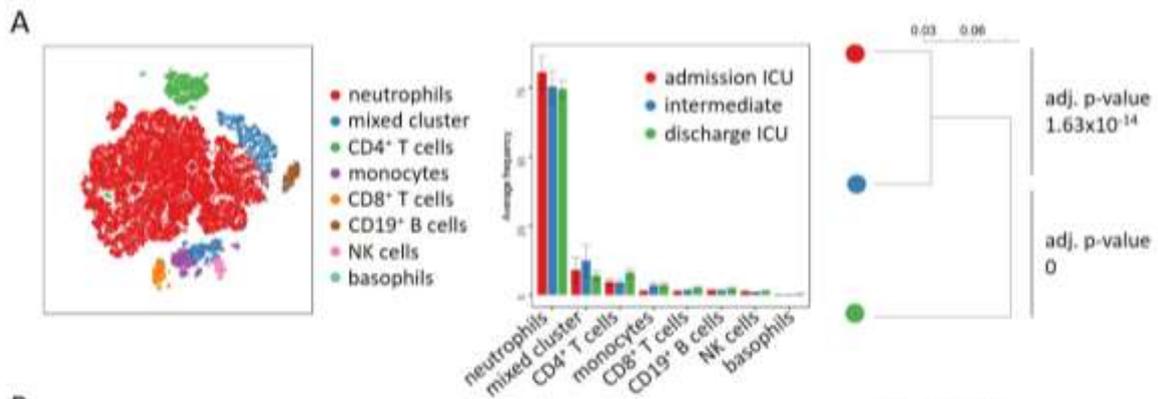
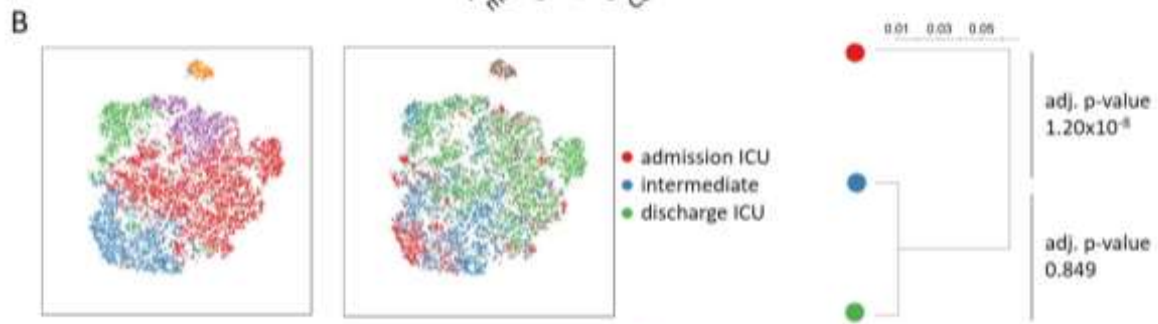
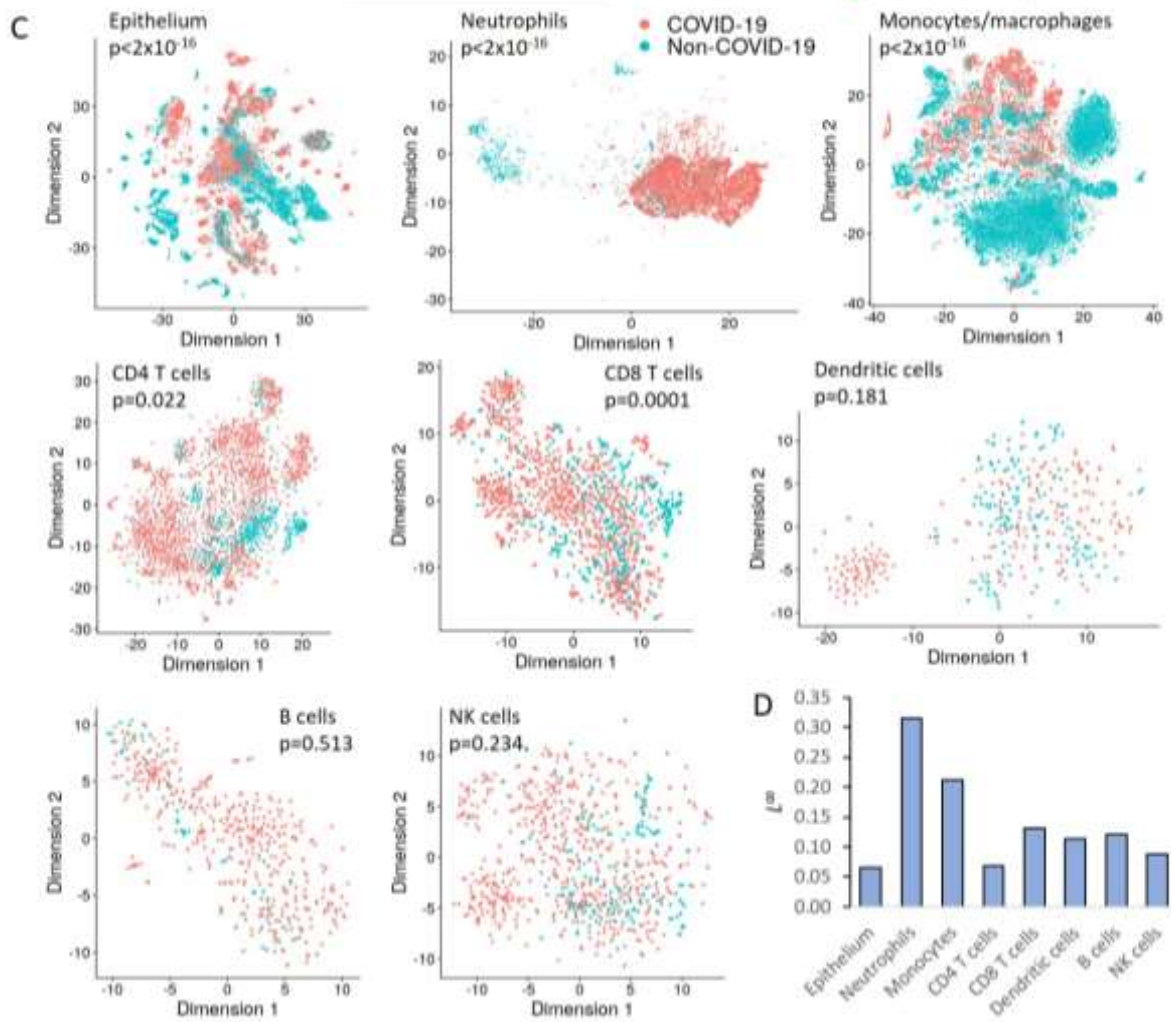

**Figure 5. A Cross Entropy test provides a robust statistical test for UMAP comparison.**
The MUS flow cytometry dataset was used to test the sensitivity of the cross entropy test. **A)** Three technical replicates of a single splenocyte sample were plotted on a UMAP with FlowSOM clustering in overlay. P values for Cross Entropy testing are shown. **B)** UMAP plots of splenocytes from three individual age/sex-matched mice (biological replicates). P values for Cross Entropy testing are shown. **C)** UMAP plots from lymph nodes, spleen and tissue (small intestinal lamina propria). P values for Cross Entropy testing are shown. **D)** Two independent UMAP runs of lymph node cells. P values for Cross Entropy testing are shown. **E)** Cumulative distribution function of p-values obtained from 400 comparisons of independent UMAP runs of the same lymph node sample. Statistical comparison to uniform distribution using KS test. **F)** In order to compare samples with biological difference restricted only to qualitative or quantitative changes, we used the previously generated artificial samples where lymph node clusters were rescaled to splenic frequencies ($LN^{\%spleen}$) or splenic clusters were rescaled to the LN frequencies ($Spleen^{\%LN}$), and compared both against the parental spleen sample in UMAP. P values for Cross Entropy testing are shown.

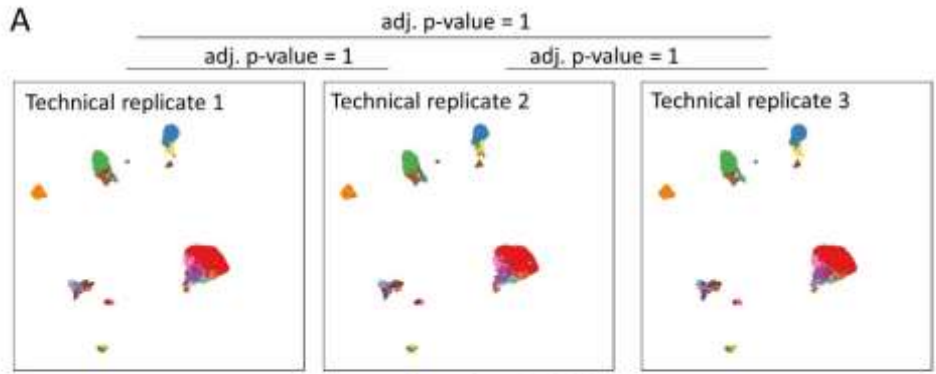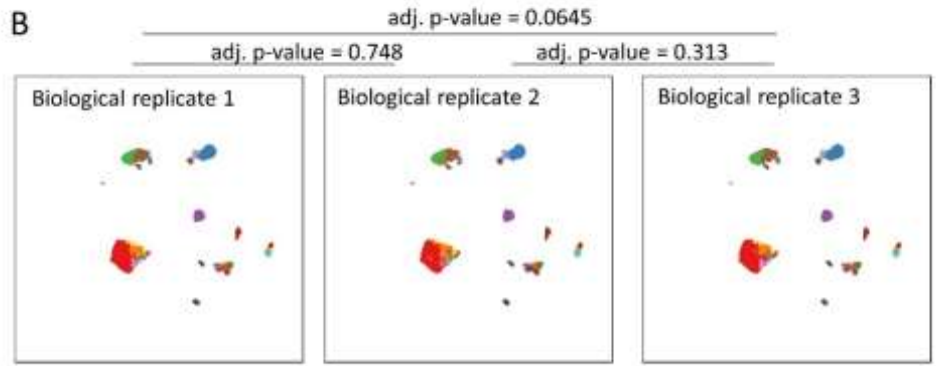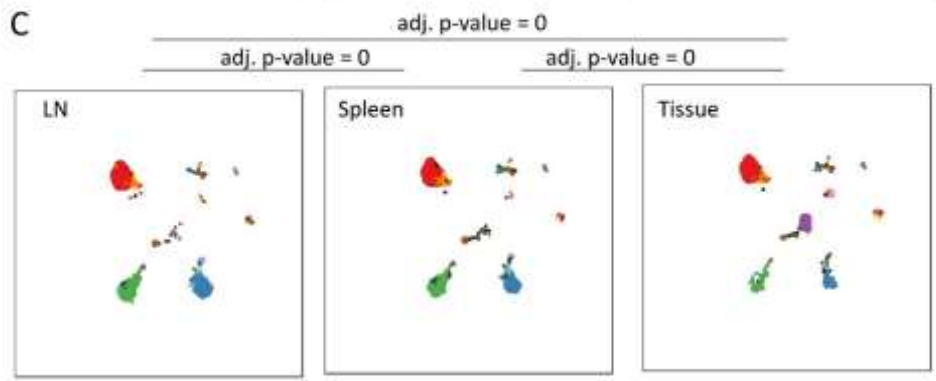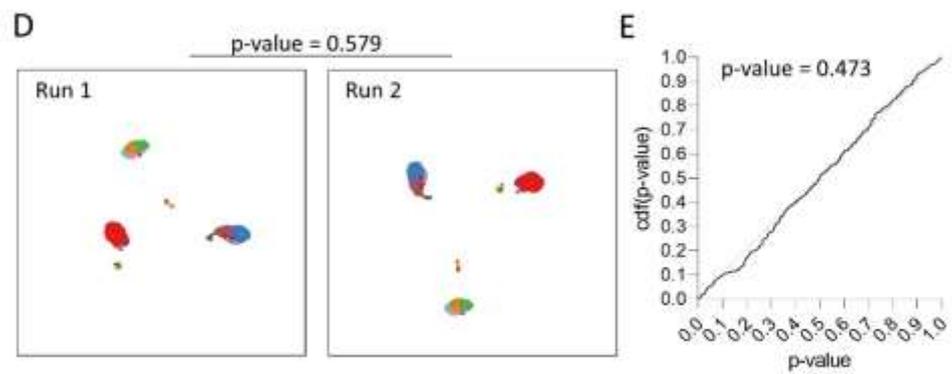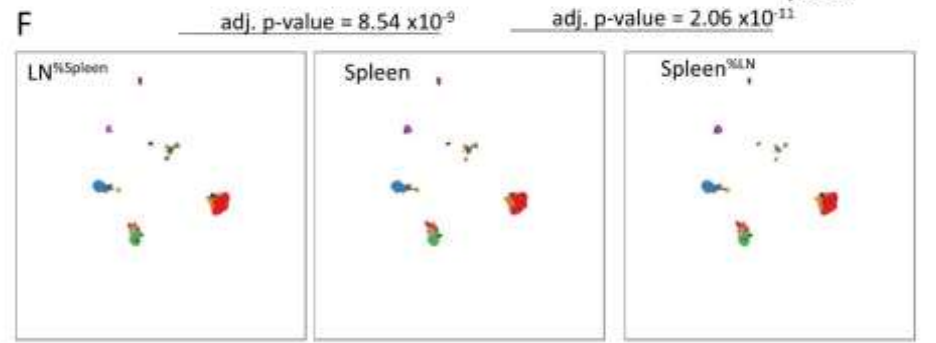